\documentclass{article}
\usepackage{amsmath}
\usepackage{mathrsfs}
\usepackage{mathtools}
\usepackage{xcolor}
\usepackage[margin=1in]{geometry} 
\usepackage{authblk}
\usepackage{fontawesome}

\title{Canonical Energy-Momentum Tensor of Abelian Fields}
\author[1,\faEnvelopeO ]{Peir-Ru Wang}
\author[1,*]{Yen-Cheng Chang}

\affil[1]{Department of Physics, National Tsing Hua University, Hsinchu, Taiwan}
\affil[\faEnvelopeO ]{louisnthu@gapp.nthu.edu.tw}
\affil[*]{ygoultraman@gapp.nthu.edu.tw}
\date{\today}

\begin{document}

\maketitle

\begin{abstract}
In this tutorial, we provide the natural derivation of symmetrical, gauge-invariant canonical energy-momentum tensor for the abelian gauge field, i.e., the electromagnetic field.
\end{abstract}

\section*{Introduction}

The energy momentum tensor should be symmetric due to the conservation of angular momentum [1].  For a given Lagrangian $\mathscr{L}$ of a field, there are two major ways to obtain the energy momentum tensor: from general relativity or from the canonical way. The energy momentum tensor obtained from general relativity is automatically symmetric. However, the canonical energy momentum is not symmetric, in general, if the field is not a scalar field. This situation occurs in an electromagnetic field. This asymmetric issue also causes the canonical energy-momentum tensor not to be gauge-invariant. Several ways are used to fix this asymmetric and resolve the gauge invariant, but all seem unnatural [2].

\section*{Derivation}

We denote $p$ is a point in spacetime, $A(p)$ as gauge potential 1-form of the electromagnetic field, and the field strength is $F = dA$. We will discuss the effect of the variation on gauge potential and the spacetime variation. We denote the variation on gauge potential,
\[ A \rightarrow A + \delta A \]
, and the spacetime variation drag by a vector field $\delta x$ denote as:
\[ p \rightarrow \tilde{p} = f_{\delta x}(p) \]
The total variation of gauge 1-form is
\[ \Delta A = \tilde{A}(\tilde{p}) - A(p) = \delta A + \hat{\mathcal{L}}_{\delta x} A \]
, where $\hat{\mathcal{L}}_{\delta x} A$ is the Lie derivative of $A$ with respect to the vector field $\delta x$, instead of $\Delta A=\delta A+\delta x^\mu\partial_\mu A$ in traditional calculation.
In local coordinate $p \rightarrow \{x^\gamma\}$, the expressions are
\begin{align*}
A(p) &\rightarrow A_\nu(x^\gamma) 
\end{align*}
\begin{align*}
F &\rightarrow F_{\mu\nu} = \partial_\mu A_\nu - \partial_\nu A_\mu 
\end{align*}
\begin{align*}
p \rightarrow \tilde{p} &\rightarrow  \tilde{x}^\gamma = x^\gamma + \delta x^\gamma
\end{align*}
\begin{align*}
    \Delta A_\nu &= \tilde{A}_\nu(\tilde{x}) - A_\nu(x) = \delta A_\nu + A_{\nu,\gamma}\delta x^\gamma + A_\gamma \delta x^{\gamma}_{,\nu}
\end{align*}

The Lagrangian $\mathscr{L}$ and the action $S$ of electromagnetic field are
\begin{align*}
 \mathscr{L} = -\frac{1}{16\pi c} g^{\mu\alpha}g^{\nu\beta}F_{\alpha\beta}F_{\mu\nu}\sqrt{-g} 
\end{align*}
and
\[ S = \int d^4x \, \mathscr{L}[A_\mu(x^\gamma), A_{\nu,\mu}(x^\gamma), x^\gamma] \]
We derive the equation of motion (EoM) and Noether theorem follow standard procedure. The variation of action $\Delta S$ divides into two terms:
\[ \Delta S = \int  \Delta d^4x \cdot \mathscr{L} + \int d^4x \cdot \Delta \mathscr{L} \]
The first term is the variation of volume form, which is [3]
\[ \Delta d^4x =  \delta{x}^\gamma _{,\gamma} \cdot d^4x \]

\noindent The second term is the variation of $\mathscr{L}$

\begin{align*}
\Delta \mathscr{L} &= \mathscr{L}[\tilde{A}_\nu(\tilde{x}^\gamma), \tilde{A}_{\nu,\mu}(\tilde{x}^\gamma), \tilde{x}^\gamma] - \mathscr{L}[A_\nu(x^\gamma), A_{\nu,\mu}(x^\gamma), x^\gamma] \\
& = \left[ \frac{\partial \mathscr{L}}{\partial A_\nu}  \delta A_\nu + \frac{\partial \mathscr{L}}{\partial (\partial_\mu A_\nu)}  \delta (\partial_\mu A_\nu) \right] (x^\gamma)  + \left[ \mathscr{L}_{, \gamma} \delta x^\gamma \right] (x^\gamma)  + O(\delta^2)
\end{align*}

\noindent Hence, $\Delta S$ is

\begin{align*}
    \Delta S &= \int d^4x \cdot \underbracket[0.4pt][0pt]{\mathscr{L} \delta x^\gamma_{,\gamma}}_{(A)} + \int d^4x \left[ \frac{\partial \mathscr{L}}{\partial A_\nu}\delta A_\nu + \frac{\partial \mathscr{L}}{\partial(\partial_\mu A_\nu)}\delta(\partial_\mu A_\nu) + \underbracket[0.4pt][0pt]{\mathscr{L}_{,\gamma} \delta x^\gamma}_{(B)} \right] \\
    &= \int d^4 x \cdot \left[ \frac{\partial \mathscr{L}}{\partial A_\nu}\delta A_\nu + \frac{\partial \mathscr{L}}{\partial(\partial_\mu A_\nu)}\partial_\mu(\delta A_\nu) + \underbracket[0.4pt][0pt]{(\mathscr{L} \delta x^\gamma)_{,\gamma}}_{(A)+(B)} \right]\\
    &= \int  \left[ \frac{\partial \mathscr{L}}{\partial A_\nu} - \partial_\mu\left(\frac{\partial\mathscr{L}}{\partial(\partial_\mu A_\nu)}\right) \right] \delta A_\nu d^4 x +  \int \left[ \partial_\mu\left(\frac{\partial \mathscr{L}}{\partial (\partial_\mu A_\nu)}\delta A_\nu\right) + (\mathscr{L} \delta x^\gamma)_{,\gamma}\right] d^4 x 
\end{align*}
The EoM is
\[
\frac{\partial \mathscr{L}}{\partial A_\nu} - \partial_\mu \left( \frac{\partial \mathscr{L}}{\partial(\partial_\mu A_\nu)} \right) = 0
\]
\noindent Now we introduce the total variation $\Delta A = \delta A + \hat{\mathcal{L}}_{\delta x} A $, instead of the traditional $\Delta A_\nu=\delta A_\nu +\delta x^\mu \partial_\mu A_\nu$ (see appendix).

\noindent Using $\delta A_\nu = \Delta A_\nu - A_{\nu,\gamma} \delta x^\gamma - A_\gamma \delta x^\gamma_{,\nu}$,
\begin{align*}
    \Delta S &=  \int \{EoM\} \delta A_\nu d^4 x + \int \partial_\mu\left[\frac{\partial \mathscr{L}}{\partial(\partial_\mu A_\nu)} (\Delta A_\nu - A_{\nu,\gamma}\delta x^\gamma - A_\gamma \delta x_{,\nu}^\gamma) + \delta^\mu_\gamma\mathscr{L} \delta x^\gamma\right]d^4x\\
     &=  \int \{EoM\} \delta A_\nu d^4 x + \int \partial_\mu \left[ \frac{\partial \mathscr{L}}{\partial(\partial_\mu A_\nu)}\Delta A_\nu -  \left( \frac{\partial \mathscr{L}}{\partial(\partial_\mu A_\nu)}A_{\nu,\gamma} \delta x^\gamma + \underbracket[0.4pt][0pt]{\frac{\partial \mathscr{L}}{\partial(\partial_\mu A_\nu)}A_\gamma \delta x^\gamma_{,\nu}}_{(*)} - \delta^\mu_\gamma \mathscr{L}\delta x^\gamma\right)\right]d^4x
\end{align*}

\noindent Evaluate the $(*)$ term:

\begin{align*}
\underbracket[0.4pt][0pt]{\partial_\mu \left[ \frac{\partial \mathscr{L}}{\partial(\partial_\mu A_\nu)} A_{\gamma} \delta x^\gamma_{,\nu} \right]}_{(*)}
&= \underbracket[0.4pt][0pt]{\left[ \frac{\partial \mathscr{L}}{\partial (\partial_\mu A_\nu)}  A_\gamma \delta x^\gamma \right]_{,\nu \mu}}_{(1)} - \underbracket[0.4pt][0pt]{\partial_\mu \left[ \left( \frac{\partial \mathscr{L}}{ \partial (\partial_\mu A_\nu)}\right)_{,\nu}  A_\gamma \delta x^\gamma\right]}_{(2)} - \underbracket[0.4pt][0pt]{\partial_\mu \left[\frac{\partial \mathscr{L}}{\partial(\partial_\mu A_\nu)}A_{\gamma,\nu} \delta x^\gamma \right]}_{(3)}
\end{align*}
Note that
$\frac{\partial\mathscr{L}}{\partial(\partial_\mu A_{\nu})}$ is

\begin{align*}
\frac{\partial \mathscr{L}}{\partial(\partial_\mu A_\nu)}
    &= -\frac{1}{4\pi c} g^{\mu\alpha}g^{\nu\beta} \sqrt{-g}  F_{\alpha\beta} 
\end{align*}

\noindent The $(1)$ term
\[\left[ \frac{\partial \mathscr{L}}{\partial (\partial_\mu A_\nu)} \right]_{,\nu \mu} A_\gamma \delta x^\gamma = \left[ -\frac{1}{4\pi c} g^{\mu\alpha}g^{\nu\beta} \sqrt{-g}  F_{\alpha\beta} \right]_{,\nu \mu} A_\gamma \delta x^\gamma  = 0\]
due to the antisymmetric of $F_{\mu\nu}$ and the symmetric of second order derivative $ \{_{,\nu\mu}\}$.

\noindent Since $\frac{\partial \mathscr{L}}{\partial A_\nu} = 0$, the EoM becomes

\[ \partial_\mu \left( \frac{\partial \mathscr{L}}{\partial (\partial_\mu A_\nu)} \right) = \left( -\frac{1}{4\pi c} g^{\mu\alpha}g^{\nu\beta}F_{\alpha\beta}\sqrt{-g} \right)_{,\mu}= 0 \left( =\frac{\partial \mathscr{L}}{\partial A_\nu} \right)\]
Then the $(2)$ term:
\[ \left( \frac{\partial \mathscr{L}}{\partial (\partial_\mu A_\nu)} \right)_{,\nu} = \left( -\frac{1}{4\pi c} g^{\mu\alpha}g^{\nu\beta}F_{\alpha\beta}\sqrt{-g} \right)_{,\nu} =  \left( \frac{1}{4\pi c} g^{\mu\alpha}g^{\nu\beta}F_{\beta\alpha}\sqrt{-g} \right)_{,\nu} = 0 \]
Hence $(*)$ only left $(3)$ term:
\[
\underbracket[0.4pt][0pt]{\partial_\mu\left[\frac{\partial \mathscr{L}}{\partial(\partial_\mu A_\nu)} A_\gamma \delta x^\gamma \right]_{,\nu}}_{(*)} = \underbracket[0.4pt][0pt]{\partial_\mu\left[\frac{\partial \mathscr{L}}{\partial(\partial_\mu A_\nu)} A_{\gamma, \nu} \delta x^\gamma \right]}_{(3)}
\]
The final result,
\begin{align*}
\Delta S &= \int \{EoM\}\delta A_\nu d^4 x + \int\partial_\mu \left[ \frac{\partial \mathscr{L}}{\partial(\partial_\mu A_\nu)}\Delta A_\nu -  \left( \frac{\partial \mathscr{L}}{\partial(\partial_\mu A_\nu)} A_{\nu,\gamma} \delta x^\gamma -  \underbracket[0.4pt][0pt]{ \frac{\partial \mathscr{L}}{\partial(\partial_\mu A_\nu)} A_{\gamma, \nu}   \delta x^\gamma }_{(*)=(3)}  \right) - \delta^\mu_\gamma \mathscr{L}\delta x^\gamma \right]d^4x \\
&= \int \{EoM\}\delta A_\nu d^4 x + \int \partial_\mu \left[ \frac{\partial \mathscr{L}}{\partial(\partial_\mu A_\nu)}\Delta A_\nu -  \left( \frac{\partial \mathscr{L}}{\partial(\partial_\mu A_\nu)} F_{\gamma\nu} - \delta^\mu_\gamma \mathscr{L} \right) \delta x^\gamma  \right]d^4x
\end{align*}

\noindent Hence we have
\begin{align*}
T^\mu_\gamma &= \frac{\partial \mathscr{L}}{\partial(\partial_\mu A_\nu)} F_{\gamma\nu} - \delta^\mu_\gamma \mathscr{L} \\
 &= -\frac{1}{4\pi c}F^{\mu\nu}F_{\gamma\nu}\sqrt{-g} + \delta^\mu_\gamma \frac{1}{16\pi c} F^{\alpha\beta}F_{\alpha\beta}\sqrt{-g}
\end{align*}
is symmetric and gauge invariant.

\section*{Summary}

The natural symmetrical, gauge-invariant canonical energy-momentum tensor for the abelian gauge field is derived. This derivation does not depend on flat spacetime geometry, hence is background independent. This method has potential to cover the non-abelian field theory and general relativity.

\section*{References}
\begin{enumerate}
    \item L.D. Landau, The Classical Theory of Fields, Elsevier 2013.
    \item A. Freese, Noether's theorems and the energy-momentum tensor in quantum gauge theories, Physical Review D, 106 (2022) 125012.
    \item L.H. Ryder, Quantum Field Theory, Cambridge University Press, 2nd Edition, 1996.
\end{enumerate}

\end{document}